# i-Pulse: A NLP based novel approach for employee engagement in logistics organization

Rachit Garg [a,b,*], Arvind W Kiwelekar [a], Laxman D Netak [a], Akshay Ghodake [b]

[a] Department of Computer Engineering Dr. Babasaheb Ambedkar Technological University, Lonere, Dist Raigad (MS) 402103, India
[b] Center for Advanced Technologies and Innovation ATA Freight Line India Pvt. Ltd., Pune, Maharashtra 411028, India

ABSTRACT

Although most logistics and freight forwarding organizations, in one way or another, claim to have core values. The engagement of employees is a vast structure that affects almost every part of the company's core environmental values. There is little theoretical knowledge about the relationship between firms and the engagement of employees. Based on research literature, this paper aims to provide a novel approach for insight around employee engagement in a logistics organization by implementing deep natural language processing concepts. The artificial intelligence enabled solution named Intelligent Pulse (i-Pulse) can evaluate hundreds and thousands of pulse survey comments and provides the actionable insights and gist of employee feedback. i-Pulse allows the stakeholders to think in new ways in their organization, helping them to have a powerful influence on employee engagement, retention, and efficiency. This study is of corresponding interest to researchers and practitioners.

## 1. Introduction

Over the last decade, the way industry works worldwide has undergone drastic changes with practically disappearing traditional financial, political, and geographical boundaries. Such developments in global business conditions have not left the logistics sector unaffected (Bhattacharya, 2015; Winkelhaus and Grosse, 2020).In recent years, both via AI and Blockchain, the logistics industry's Community is undergoing a significant transformation. The logistics value chain is an integral part of the company's entire value chain (Batta et al., 2020). Supply chains are similar to the actual metal chain. Each step in the production and delivery of the product is a connection that plays a critical role in transforming the concept into a real, deliverable product. However, when the link weakens, the entire chain loses strength-or breaks entirely.

Logistics firms are the backbone of many other sectors, keeping things going in the right way and on time. Without an appropriate internal communication system in place, workers are at risk of feeling isolated, which can reduce efficiency and income. Disconnected and frustrated workers are one of the weakest links in the supply chain. Lean logistics[1] is about building efficiencies, which is just what employee engagement is all about. The effects of the workers engaged (decreased turnover, improved efficiency, and like) are what you would like to see in a lean logistic environment. Increasing globalization and industrialization have generated imminent green and sustainable logistics (G&SL) requirements. G&SL initiatives have generated global discussions over recent decades, seeking to reduce the external negative transport conditions and improve the supply chain's efficiency (Ren et al., 2020). An organization's culture and employees' engagement are essential for new green supply chain practices (Elzarka, 2020). Concerning the management of the increased diversity of workers, including various perceptual, cognitive, and physical skills, there is a strong need to create individualized, personalized solutions (Sgarbossa et al., 2020). Deloitte says that the average growth of three-year revenue was 2.3 times that of highly engaged employed organizations with average staff engagement (Daichendt and Kaplan, 2016). Paul J. Zak found that employees working in high trust companies enjoy their jobs 60% more, 70% more aligned with company goals, and feel 66% closer to their colleagues (Zak, 2017).

If there is a need for continuous improvement, staff need to be engaged and enthusiastic, which is what commitment is all about. When workers are engaged, they are adequately feeding into lean logistics. Benefits of employee engagement in the supply chain often improve the productivity and efficiency of the shipment chain. It is a significant factor in workers' turnover since disengaged staff members frequently feel under-equipped, undervalued, and antiquated. Logistics leaders who consider the inspiration of engagement and take steps for employee engagement would also benefit both in the long term and the short term.

---

* Corresponding author.
  *E-mail addresses:* rachit.garg.nitttr@gmail.com (R. Garg), cati_5@atafreight.com (A. Ghodake).

[1] It is a method in which unnecessary practices are identified and removed. Lean logistics improve efficiencies, generating positive customer engagement.

Leaders in the supply chain need to take steps to address the problem of engagement. However, action means taking the time to consider what causes disengagement in the first place and then trying to address the underlying issues. Although there is a need for research is on the front end, increasing employee engagement produces a better, more productive supply chain in the long run.

Studies report that only 14 to 30 percent of workers are engaging at work (Welbourne, 2007). Retaining brilliant employees in the organization is not so challenging, although keeping them engaged is the biggest challenge that companies are facing today. Management needs to win their hearts and minds during their career path. Employee engagement in the dynamic market today has proven a crucial catalyst for business success. In addition, employee participation may be a key factor in a successful logistics business. Engagement is a key factor for enhancing the credibility of a company and its stakeholders. Innovation, Organizational culture, communication barriers, relationships, and conflict resolutions are among the several parameters contributing to a logistics organization's successful procurement (Kar and Pani, 2014). All these parameters also contribute to enhancing employee engagement in an organization. Committed employees can do more, that is, to perform beyond expectations and to have a high correlation with logistics organizations' innovation behaviors. It includes an organizational culture that promotes innovation and reduces communication barriers for its employees. The more the employees are engaged, the stronger the relationship between them. In other words, the engagement of employees corresponds directly to the employee's intimacy of the relationship. Employee engagement has the ability to profoundly impact the retention and loyalty of employees in the logistics organization. Job satisfaction is often used synonymously with employee participation as both concerns the employees, workplace atmosphere, compensation, and benefits. The advantage of employee engagement in a logistics organization, including lowering absenteeism by 41%, less turnover by 24%, and 70% reduction in safety incidents, contribute to short-term and long-term gains (Post, 2020). Employee engagement is a dynamic term, with several factors influencing the degree of commitment. Consequently, there are several ways to encourage interaction, but no set that suits all organizations. As a complement to conventional annual surveys, pulse surveys are becoming popular, concentrating on topics such as employee engagement or employee satisfaction. Pulse surveys and more regular employee surveys are standard remedies to the frequent employee engagement issue found by extensive surveys. It is usually the proportion of the workforce that demonstrates a high degree of dedication to answering a particular set of questions. Technology standards are highly desirable by the companies to develop broader repositories that could integrate survey results with HR management systems. To remain competitive in the market, organizations must promote positive employee engagement as a strategic action for gaining a competitive advantage for the company.

This paper discusses the enhancement of employee engagement in a logistics enterprise by introducing a new model named i-Pulse. This article builds on the deep natural language processing system that automatically analyzes thousands of pulse survey comments to determine the keywords most important to an organization. i-Pulse takes advantage of people's data in real-time to give logistics leaders an overall snapshot of the company's well-being, insight to identify challenges and action directions. This paper is present in six sections. Section 2 discusses the theoretical model by identifying the employee engagement constructs from the literature. Section 3 discusses the motivation behind the use of NLP in logistics employee engagement. Section 4 discusses the method of implementation. Section 5 and 6 discuss the result and discussion of this research. The conclusion of the study is present in section 7 of this paper.

**2. Literature review**

Employee engagement in the last few years has been of great interest. Many argued that employees' engagement determines employee productivity, organizational progress, and financial efficiency (Saks, 2006; Shanmugam and Krishnaveni, 2012). In recent years, the commitment of employees has become a widespread issue. Nevertheless, critical academic literature on this issue remains unknown; it is relatively unknown how management can significantly affect employee engagement. As there is a growing concern in engagement, there is much uncertainty as well. At present, there is no concept of consistency, with the contribution being operationalized and measured in several different ways. Authors also carried out a literature survey by reviewing journal articles, working papers, textbooks, case studies, and other publicly accessible publications concerning employee engagement (Refer Fig. 1). Literature often interchangeably uses logistics and Supply Chain Management (SCM), but the two differ subtly. SCM is more strategic than logistics is more operational (Mitra, 2008). The review also discusses shortcomings and problems not yet addressed, indicating where further investigations should occur. The review attempts to add value to the current state of understanding by a thorough examination of established employee engagement literature and reflecting on existing discussions and conclusions.

The lack of a universally acceptable definition of employee engagement is one of the initial challenges found in the literature. Provided the lack of common understanding around the sense of employee engagement, to establish the construct, it is highly fragmented. The term "employee engagement" is not specified precisely (Wahba, 2015). Macey et al. stated the ambiguity in the academic and practical implementation of concepts of employee engagement. They also introduced the confusion between the psychological states of employee engagement presented by various researchers. They also discussed the 14 Prepositions related to psychological state engagement, behavioral engagement, and trait engagement (Macey and Schneider, 2008). According to Dalal et al., the literature on employee engagement is in a state of disarray. They also updated Macey and Schneider's hypothesis by clarifying how the psychological state of engagement and word engagement is not only cognitive-affective but also dispositional and behavioral (Dalal et al., 2008).

**The beginning: from paper to web**

Kahn introduced the earliest concept of employee participation in scholarly literature in 1990. Kahn says that engagement means that employees are present both psychologically and physically in the organizational role (Kahn, 1990). According to Truss et al. in 2006, employee engagement is a passion for work. To realize and compute employee engagement, they identified five main components to be analyzed, and those are Overall engagement, Cognitive engagement, Emotional engagement, Physical Engagement, and Advocacy (Bailey et al., 2020). The definition of employee work passion was introduced in another study by Zigarmi et al.; employee work passion says that work passion derives from the regular cognitive and emotional appraisal of different work and organizational circumstances. They also introduced a work passion model that integrates the idea of appraisal into employee work passion. They agree that the three components of cognition, affect, and intention must be integrated into every useful concept of employee passion, dedication, or satisfaction. It can be achieved by applying the appraisal construct and the elements of cognition, affect, and intention (Zigarmi et al., 2009). Engagement is a positive and satisfactory mindset linked to work with a strong sense of strength, dedication, and absorption (Schaufeli and Bakker, 2004). Employees' sentimental and mental commitment to the organization defines the employee engagement or affective commitment of employees at work (Baumruk, 2004; Richman, 2006; Shaw, 2005).

The fact that there are different interpretations makes it challenging to assess the degree of understanding of employee engagement as each research explores the participation of employees under a separate framework. It is challenging to measure the engagement until it is universally defined (Ferguson and Carstairs, 2005). According to

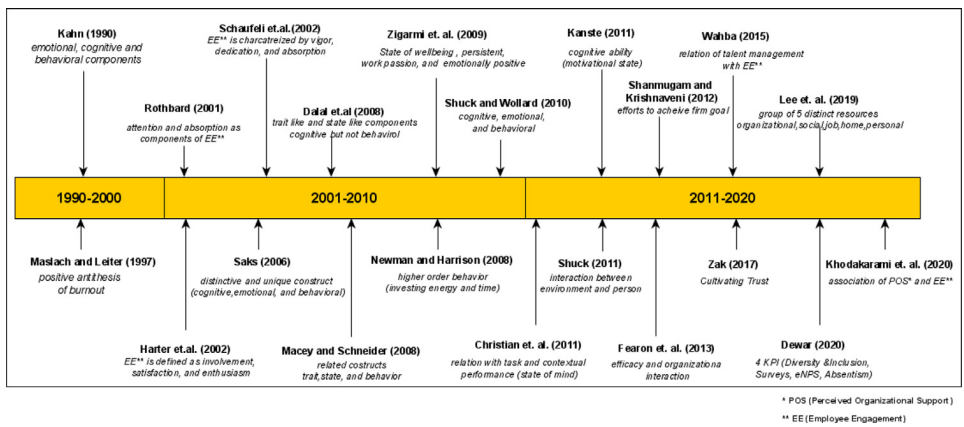

Fig. 1. Timeline of growth in employee engagement concepts.

Harter et al. (2002), there is a decisive and fascinating link between employee engagement and business outcomes. These include employee performance and efficiency, security, productivity, presence, maintenance, profitability, customer service, and customer satisfaction. According to Newman, employee engagement is the investment of personal resources like time and energy. They also stated that the positive diversity in organizational support activities might be explicitly derived from engagement (Newman and Harrison, 2008). The shipping sector has lacked professional and experienced personnel over the past two decades and is facing a severe staff retention problem (Haka et al., 2011; Lewarn and Francis, 2009). In the maritime industry, there is a relatively high mobility of staff between the various shipping industries (Fei et al., 2009). The degree of movement among the employees switching from one employer to another is high. The financial impact of this switching in the logistics industry can be high as the organization has already spent the efforts, time, and money on the departing employee. According to Mason (2019), the interrelation between management, employee engagement, and loyalty in a logistics service company is the cornerstone.

Measuring employee engagement is a challenge in an organization. Many logistics companies use the standard approach of the annual employee survey as the measuring tool for getting insights. In order to assess patterns over a period of time, surveys can often be beneficial. Many practitioners and consultancies have developed their survey methods and definitions for measuring employee engagement (Christian et al., 2011; Kanste, 2011). One of the best-known measurement methods is Gallup Workplace Audit (GWA), developed by the Gallup organization. The methods are build from inspired research, satisfaction surveys, group productivity, and supervisory experiences (Harter et al., 2003; O'Boyle and Harter, 2013). Shuck et al. systematic review of 1009 sources examines the wide range of literature on employee engagement. They offered an overview of the scholarship concept. This research also provides guidelines for practitioners in determining the extent of engagement (Shuck and Wollard, 2010). He also defined four essential methods to measure engagement actions: (1) the Burnout Antithesis Approach; (2) the Need-Satisfying Approach; (3) the Satisfaction Engagement Approach; (4) The Multidimensional Approach (Fletcher and Robinson, 2013; Shuck, 2011).

(1) **Measures from the Burnout-Antithesis Approach**
The positive and negative facets of a coin are engagement and burnout. It will eventually lead people who are more engaged and are eventually low on burnout and vice versa (Crawford et al., 2010; Maslach et al., 2001). Commitment includes energy, engagement, and productivity, while fatigue, frustration, and lack of success are opposite factors that explain burnout. The steps taken from the burnout-antithesis method tend to focus on activated energy sources, which are demonstrated by the adjectives used, e.g., bursting, strength, motivating, vigilant, alertness.

(2) **Measures from the Needs-Satisfying Approach**
Measures developed through the need-satisfactory method derive directly from Kahn's theoretical propositions and empirical findings "psychological engagement" research. Kahn proposed that a person can be physically active, engaged emotionally, and engage in cognitive activity (Kahn, 1990). Three psychological dimensions: meaningfulness, stability, and availability greatly influence these states of life. Maslach et al. identify dedication as positive burnout antethesis. They also established the association between the three main aspects of burnout and six fields of organizational life (Maslach and Leiter, 1997).

(3) **Measures from the Satisfaction-Engagement Approach** The measures from this approach primarily aim to improve workplace or organizational life (Harter et al., 2002). Therefore, this approach is for the particularity of managerial practice and organizational initiatives (MacLeod and Clarke, 2011).

(4) **Measures from the Multidimensional Approach**
The multidimensional method is the latest breakthrough in the field of employee engagement and originated from Saks's research (Saks, 2006). This approach is somewhat similar to the need-satisfaction approach since it concentrates on role performance. However, it is distinct because it distinguishes between the focus of the job and the organization.

Schaufeli et al. initiated the conceptualization of engagement by describing well-being linked to work. They also regarded burnout and engagement as opposing terms. Burnout is defined by a mixture of exhaustion (low activation) and cynicism (low identification), while engagement is characterized by vigor (high activation) and dedication (high identification) (Schaufeli et al., 2002). Johnson (Johnson, 2004) defines that implementing specific engagement rules will bring employee engagement and make the organization effective and profitable. Emotional intelligence brings an emotional economy to the organization. According to him, employing rules in engagement costs nothing to a company but builds bonding between the company and employees. Another separately designed survey that measures the engagement benchmarks and evaluates its context is by a professional organization named Towers Perrin (Perrin, 2003). Their survey includes the blend of tangible and intangible elements and thus tracks the views and attitude of employees to understand the elements of work experience that drive attraction, retention and, engagement. According to a professional firm Prince Waterhouse Coopers (PWC), the definition of employee engagement is the ability of an employee who can act and exercise flexibility to achieve business results. Development Dimensions International (DDI) also implemented an employee engagement survey (Wellins et al., 2005). The items of the DDI engagement survey illustrate the approach to satisfaction. Many of the items in this survey measure job satisfaction precisely, and others measure the sub-dimension of work satisfaction. Rothbard introduced a nine-item scale to measure the two components of engagement attention and absorption (Rothbard, 2001). In a recent study by Lee et al. in 2019, they presented a taxonomy of resources from macro to micro-level required to influence engagement by revising 137 articles. These resources have technical, fiscal, or physical features, psychological, emotional, or cognitive features, conditions, or energy. The author categorized these resources into five distinct categories those are (a) organizational resources, (b) social resources, (c) job resources, (d) home resources, and (e) personal resources. (Lee et al., 2019). In addition to Rothbard (2001) research, a recent study by Khodakarami and Dirani (2020) on employee engagement also states that the degree of participation varies depending on the job sector and often by gender. The results show that women are more committed than men based on their commitment. Compared to other working classes, technical and qualified employees are more engaged. This study also indicates that the degree of participation decreases proportionally due to a reduction in Perceived Organizational Support. The literature also defines the following ways of evaluating employee engagement.

### 2.1. One-on-one

One perfect way of evaluating engagement is to meet employees one-to-one. Once they know their importance, employees get the highest satisfaction. Employees who feel that their employers are genuinely concerned about their growth are more likely to be engaged. For an employer, it is necessary to create a specialized workplace experience for each team member (Chandani et al., 2016; Shuck and Wollard, 2008). An excellent way to get a real sense of what is happening with employees is the regular one-to-one meetings and informal chats with each team member. These meetings give feedback and help to create such an experience. One-On-One has various benefits such as improved interactions, identifying future problems, performance reviewing, and diagnostics (Mishra et al., 2014). A regular schedule for one-on-one logistics team meetings leads to accomplishing a consistent target and agenda.

### 2.2. Stay/Exit Interviews

The ideal way to collect input is through a formal interview with employees to figure out what encourages employees to be engaged and what hinders them from being engaged. In most organizations, exit interviews are relatively popular, but stay interviews should be a brilliant idea most often when managers ask their employees who are already satisfied at work and what makes them want to stay. The interviews at the end may be excellent, but the only problem with them is that the procedure may be too late. Exit interviewing is a popular way of knowing the reason, and it provides immediate insight into why an employee leaves the organization. These interviews can also be helpful to an organization in the retention of a valued employee (Reynolds, 2017).

### 2.3. eNPS

The concept of Net Promoter Score (NPS) for employees in various organizations is there since the last decade. The logistics and supply chain leaders have also start looking for this idea to improve employee engagement. One of the simplest and even most effective ways to evaluate engagement is by using the employee's Net Promoter Score (e-NPS) (Sedlak, 2020). For measuring loyalty, organizations use this score. The e-NPS is explicitly related to the employee's short-term intentions (Yaneva et al., 2018). It is a metric to determine employee satisfaction as part of more significant analysis. Logistics leaders seek to provide non-negotiable customer service. Considering the basis of the study conducted by Fearon et al. (2013), it is believed that constructive and consistent communication can mitigate many of the issues in logistics. e-NPS help assesses stakeholder loyalty in internal procurement and helps maintain a healthy relationship between suppliers.

### 2.4. Absenteeism rate

Another primary metric of employee engagement measurement is by measuring employee absenteeism rate. This rate gives the measure of the number of employees absent in a certain period. The lower the rate means, the higher the engagement is. Knowing the patterns in the employee absence rate may be a good indicator of employee engagement (Soane et al., 2013). However, it is a lag indicator because it is directly related to its advantage part and treats every employee in the same manner. For unavoidable circumstances like an employee with a disability, employees undergoing cancer treatment are probably to have more short-term absence (Dewar, 2020).

**Today: pulse surveys are in trend**

### 2.5. Pulse survey

A pulse survey is an emerging way of measuring the real-time vibes of a logistics organization. These short, frequent surveys are beneficial for monitoring the status of employee engagement. The pulse survey includes one or two questions to be asked to employees more frequently, fortnightly, or monthly. Pulse surveys are the perfect way to keep the office pulse constant (Lockwood, 2007). Many logistics leaders are switching to a more prominent pulse survey from the traditional way of employee engagement (Welbourne, 2016). Pulse surveys are becoming the preferred mode of engagement monitoring over a conventional annual survey that focuses on employee engagement in a logistics organization. Organizations are continually making an effort to attain the elusive goal of employee engagement. Given various attempts, the literature reveals that only there is approximately one-third of employees engagement at once. Experts say that significant changes in engagement scores could take years to come (Burjek, 2020). Literature also reveals that each established measure is dependent on the conceptualization and interpretation of the definition by independent researchers through various academic measures of employee engagement. This heterogeneity poses strong challenges in determining the importance and validity of participation studies. Many logistics leaders are also struggling with finding actionable insights from the sheer volume of information collected from weekly or monthly pulse surveys to improve employee engagement in a logistics organization. To help logistics leaders in identifying the radical cause of disengagement in an organization and breakthrough to eliminate it, there is a need for a real-time engagement measurement approach that offers an even more in-depth and more accurate picture of engagement conditions. This approach requires a framework to act on the pulse survey results system and complement the existing pulse survey measurement (Writer and Mallick, 2019).

In the last few years, pulse surveys have been prevalent in employee engagement measurement. Literature also shows many companies today looking for the next thing to happen in the engagement area. The choice of appropriate technology to enable greater employee engagement thus provides stronger links. This work aims to develop a new measuring approach, including items that cover each component of the definition proposed. It also provides actionable insights to summarize and provide a sense of employee feedback. The AI-enabled solution can evaluate hundreds and thousands of employee comments from the pulse survey to quickly summarize and provide a gist of employee feedback by identifying the most relevant keywords and phrases in a logistics organization.

### 2.6. Research objective and significance

The goal of the study is to examine the implementation of natural language processing in a logistics organization to identify insights from unstructured pulse survey comments data. To allow the supply chain leaders to take appropriate actions to balance trust and engagement and thus create an engaged workforce.

The engagement of employees has gained considerable recognition from the group of practitioners and researchers. Organizations that are dedicated to developing an atmosphere benefit enormously from employee productivity, the attainment of corporate goals, customer loyalty, and the preservation of talent. In order to understand the gist of employee feedback, it appears useful to develop the feedback analysis model and taking additional measures to build a better working atmosphere for everyone, particularly in the logistics and supply chain sector. This study will identify the quantifiable insights from pulse survey comments in a logistics organization that influence logistics leaders, and hence they can better the scope of employee engagement. From an academic viewpoint, the study contributes to NLP's potential in identifying insights from unstructured data of comments in the logistics industry.

## 3. The motivation behind NLP in logistics employee engagement

Artificial Intelligence(AI) and its subset are no more science fiction nowadays; instead, it is a digital future for many businesses. AI-based solutions are in use today to create a prosperous future. AI technology is no longer the field for futurologists but an integral part of any organization's business models and a crucial strategic factor in many business sector plans (Verma et al., 2021). AI's ability to transcend some of the computationally intensive, analytical, and maybe even innovative limitations of humans (Grover et al., 2020). With the effects on efficiency and performance, AI opens new fields of application in almost every area (Dwivedi et al., 2019). Organizations, especially for people analytic, are on the verge of adopting AI-based solutions to simplify and redefine all of their primary goals. Parthasarathy et al. have also suggested a framework using AI to boost the Cognitive Employee Management System (CEMS) performance. Their studies indicate that smart assistants based on AI can increase employee efficiency by up to 45% (Parthasarathy and Kar, 2020).

In short, artificial intelligence is a broadly applied technology term that uses algorithmic computation to transform a vast amount of data into actionable insight. The unanswered question is how to make use of technology in organizations to achieve a higher rate of employee engagement? Organizations have data collected through feedback and comments from the employees. Can practitioners in this area help the organizations to get actionable insights from this data? The employee's comments are generally lengthy and are confusing with grammatical errors, spelling mistakes, and slang. The high volume of these comments poses a considerable challenge for practitioners. Only a few organizations can spend tens (if not hundreds) of hours reading these comments manually and sorting them into quantifiable insights. For those who lack this privilege, it is an invincible challenge to translate this feedback into a meaningful data-set.

This need gap is where innovations from Natural Language Processing (NLP) and Machine Learning (ML) come into play. The ML and NLP algorithms can quickly transform the vast volumes of unstructured text data into actionable insights. Text analytics is an artificial intelligence technology that transforms unstructured data into structured information through NLP to strengthen analytics using machine learning algorithms (Kumar et al., 2021). These emerging techniques can provide new insights into employee emotions, behavior, and predictive behavior when applied to data that includes vast quantities of written or unstructured information. This combination of two key AI areas determines which employee trends are unhappy with which different percentages subgroups of employees are unhappy. This method saves time and money: it also eliminates the possibility of analysts incorporating bias in the way they categorize and interpret comments. NLP uses text analytic to provide detailed insight into employee emotions, identify problem areas, carry out thorough feedback analysis, and evaluate the survey. Using these insights, logistics leaders may quantify and develop a commitment strategy for employee engagement that discusses and encourages employee participation. A study by Chintalapudi et al. also shows the potential of text mining and text analytics in sentiment analysis on seafarers in a shipping domain (Chintalapudi et al., 2021). Besides, this integration of NLP with other data-sets or even other algorithms holds tremendous organizational potentials. The company may incorporate quantitative and qualitative survey data to create accurate opinion forecasts or even create tailored, analysis-based engagement action plans for logistics leaders. Based on a study (Teniwut and Hasyim, 2020), one of the most useful features of the current trend that can support the significant concern in the supply chain is the system that helps logistics leaders make decisions. We proposed a new approach in this study called i-Pulse that allows the logistics leaders to know which part of the employee journey would be the next priority. This approach automatically discovers patterns, issues, and opportunities from open comment text data without manual tagging.

- **i-Pulse** is an AI-enabled solution that can evaluate employee feedback comments from a pulse survey to quickly summarize and provide the essence of employee feedback by identifying the most relevant keywords and phrases in a logistics organization. It also monitors the key topics over time.
- **i-Pulse** understands the context from the comments and allows the logistics leaders to understand what their employees are talking about in a quickly manner.
- **i-Pulse** efficiently analyses text comments and obtains hands-on insights to enable the organization to respond to the most significant problems and opportunities.
- Natural language processing allows **i-Pulse** to discover patterns and trends in pulse survey comment text. Trending keywords and phrases are then automatically brought to the logistics leaders notice for more robust action plans on employee engagement.

## 4. Method of implementation

The work done in this paper is implemented in the Python. The experiment environment consists of Intel Core i5-4590 CPU, memory is 4 GB and operating system is Windows 7 professional. For the implementation of i-Pulse, the pulse comment text data is collected from global presence of ATA Freight. Fig. 4 depicts the pictorial representation of steps in implementation. The details of steps in implementation are:

(1) **Data Collection:**
   This experiment uses the dataset of employee's pulse survey comments dataset from the global presence of ATA Freight collected at Center for Advanced Technologies and Innovation (CATI) in ATA Freight Line India Pvt Ltd. At CATI, we collect the data using a short survey to check employee perceptions about an issue or idea quickly. The survey is designed by including core drivers of engagement like career, recognition, wellness, work-life balance, alignment to ambition. Once the data is collected, it is not recommended to directly go from raw text to fitting a machine learning or deep learning model. First, we have to pre-process the text by passing the raw data to a series of filters implemented in python. The pre-processing of data includes removing numbers, stop words, punctuation, and lowering the case of data. Once the data is cleaned, it is available for further processing.

(2) **Comments Relevancy:**
   The cleaned data through various pre-processing techniques are now ready to be used. The calculation of the relevancy score of each comment with other comments by using the text-similarity approach. Mathematically, by calculating the cosine angle between two vectors, the closeness between the two is calculated. In the same way, the cosine angle between the comments can determine their closeness. By using the cosine similarity, the similarity between each statement is measured. Many times we all write comments differently, but contextually they sound identical; thus, the measure of similarity indicates the relevancy between comments. The proposed method used the word embedding approach for finding similarities between comments. The method implements the word2vec algorithm by using data streaming and python interfaces. The basis of word2vec is a simple concept of a neighboring word (Refer Fig. 2). The word2vec predicts the context-based term in either of the ways by using context to predict a target word (a method known as the Continuous Bag of Words, or CBOW) or by using a word to predict a target context, which is called skip-gram. If two words have an approximately similar neighbor that is having nearly the same context, then these words possibly mean very similarly or are at least related (Ozer et al., 2018). It assumes that the association between words derives the significance of a word.

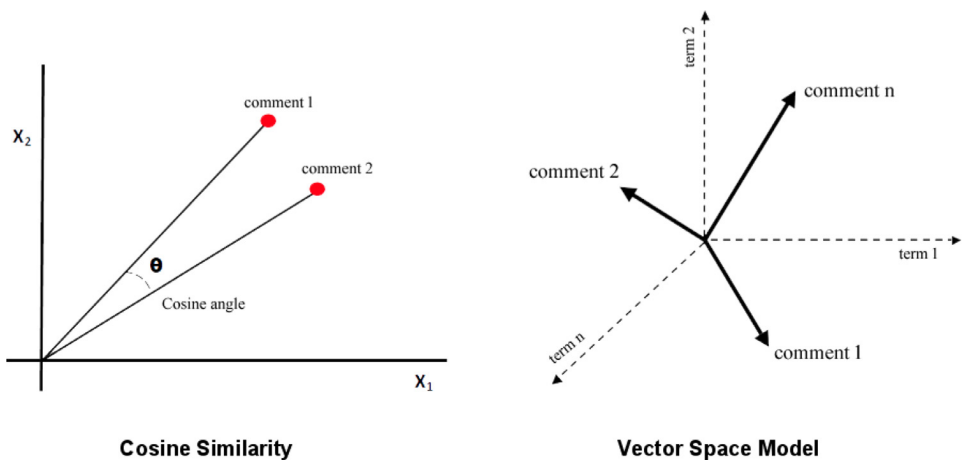

**Fig. 2.** Cosine angle and Vector Space Model of comments.

The word2vec algorithm works best when there is a massive volume of text data. A vast corpus of around 12 GB of the latest Wikipedia English articles data is used for training to find word representation in a semantic space. The word representation produced depends on how the words occur in the corpus. The latest English Wikipedia dump XML data is used for the implementation. Wikipedia is a rich source of well-structured text data sources, and it is freely and easily accessible online. It is a vast source of information that a human can expect any sort of information without giving a second thought. The downloaded Wikipedia article data needs processing from XML format Wikipedia to text format. The Python script extracts and cleans text from a Wikipedia database dump. We have been using internal multiprocessing to parallel the work and to handle the dump faster. The Wikipedia dump is downloaded on May 9, 2019, from https://dumps.wikimedia.org/. Before building the model, the plain text from the dump is extracted.

(3) **Training of Model:**
The next step is the word2vec model training from the text. A genism word2vec model file is build to train the English Wikipedia model from the Wikipedia extracted plain text. The basis of this model is word embedding. The word embedding method gives a vector space model that defines the context of each word with another word (Kenter and de Rijke, 2015; Mikolov et al., 2013). Training the model is relatively straightforward. A single hidden layer neural network to train the model for context-based prediction of the current word. The main objective here is finding the weights of the hidden layer. These weights are necessarily the word vectors that the model seeks to learn. The learned vectors obtained represent embedding. These embeddings are considered as some features that describe the target word. The algorithm used to train the word embedding is from the genism package of the python library. This algorithm is the original word2vec implementation by Google expanded with additional features. Deep learning has demonstrated promising results in the semantical representation of words (Mikolov et al., 2013). Also, for finding similarities between words, it commonly uses the word embedding approach (Kenter and de Rijke, 2015). Word embedding is trained from a broad corpus, usually linguistic or domain-specific, in order to capture the statistical connection between all words within the corpus (Nasir et al., 2021). According to Naili et al. (2017), word2vec and GloVe are better at learning the representation of vectors than Latent Semantic Analysis (LSA). Compared to Glove, word2vec presents the best word vector representation with a small dimensional semantic space. In practice, both GloVe and Word2Vec are used to translate text into embeddings, and both exhibit comparable efficiency. Although the model is trained in real applications over Wikipedia text with a corpus of around 12 GB, creating these embeddings takes a great deal of time and resources. Word2Vec conducts gradual, "sparse" neural network training by iterating over and over a training corpus repeatedly. In practice, Word2Vec uses negative samples to replace the softmax function with the sigmoid feature operating on the data to accelerate the training process. The method Term Frequency- Inverse Document Frequency (TF-IDF) focuses on the resemblance of word morphology but does not grasp inherent meanings of words. The Latent Dirichlet Allocation (LDA) method defines the similarity between subjects through word overlap. The similarity of the LDA approach is, therefore, relatively high (Liu et al., 2017). The TF-IDF maps each word with a single value and therefore does not contain any meaning, while word embedding seeks to capture the relation between words. According to Yin Luo, word2vec outperforms LDA and TF-IDF methods on the complete dataset as well as an implicit reprint of dataset (Esposito et al., 2016; Yin et al., 2018). The experiment conducted by Intellica.AI shows that the precision in text-similarity by word2vec is higher in comparison to precision achieved by TF-IDF (Intellica, 2019).

(4) **Group Formation:**
Once each comment's relevance with every other comment is calculated, the pair of relevant comments are filtered based on the threshold. The threshold considered for the filtration is 85% of the relevancy score. The low value of the threshold may give very general results, and the high value of the threshold can skip terms or phrases most relevant to each. The work represents a complex multidimensional undirected connected graph. This graph considers each comment as a node of the graph, and the edges are the relevancy between two comments. The edge between comment-1 and comment-4 means they are related to each other, with more than 85% of the relevancy score. The next move is trimming the graph into all connected components (Holberg, 1992). These sub-graphs represent the groups of relevant comments where each member (comment) of the group is relevant to each other with more than 85% of relevancy (Refer Fig. 3).

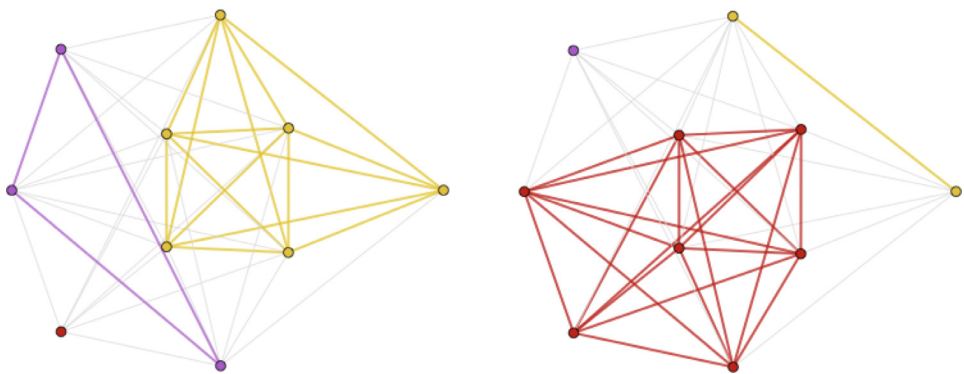

**Fig. 3.** Sub-graphs representing group formation of related comments.

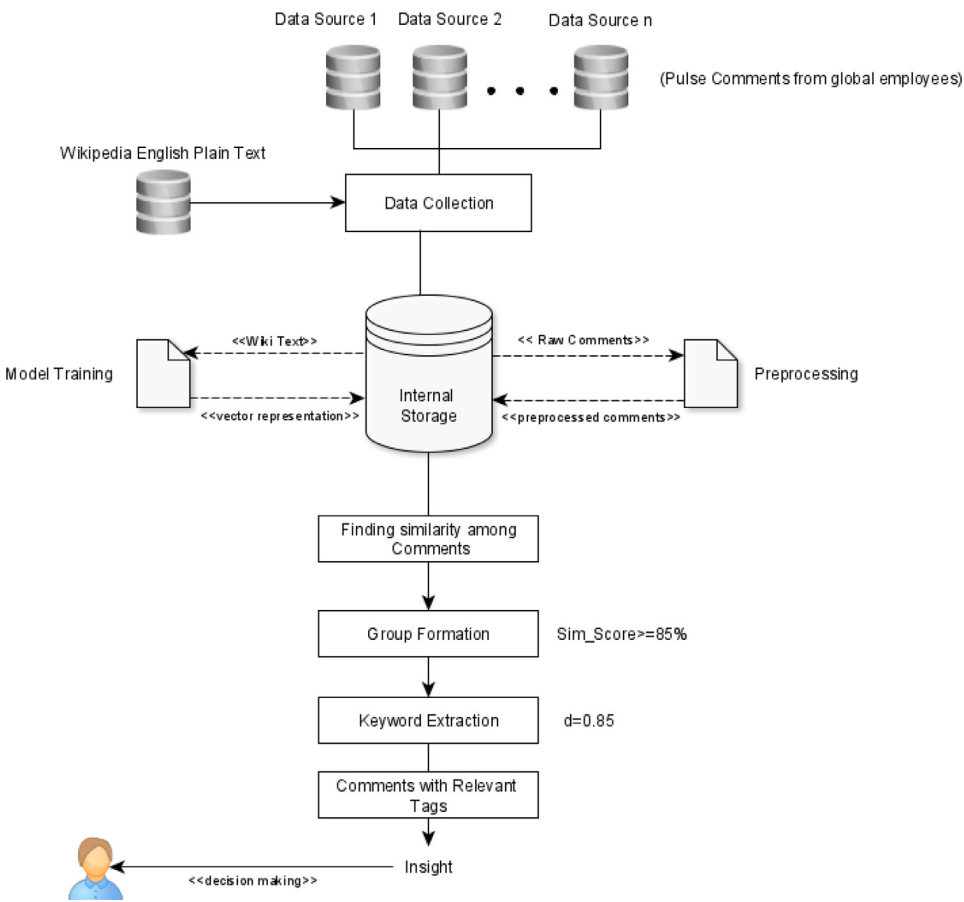

**Fig. 4.** Steps in implementation process.

The comments that are similar to each other are group together, and the resulting groups broadly describe the overall themes, topics, and patterns inside the group. The main idea behind this is to answer the questions comments in a group are entirely or in part similar if they are related to each other in terms of the same topic.

(5) **Finding Relevant Keywords from Each Group:**
Keywords describe the main topics expressed in a document. Keyword extraction is the process of automatically extracting the most relevant words and expressions from the text. It automatically detects important words that can be used to represent the text. It is a very efficient way to get insights from a vast amount of unstructured text data. Keywords are an effective way of related topic or discipline categorization of groups (Beliga, 2014). The advent of Natural Language Processing (NLP) has resulted in effective and reliable keyword extraction. For keyword extraction in this paper, we have used the TextRank algorithm (Mihalcea and Tarau, 2004) with a damping factor of 0.85 in the python environment. TextRank is a graph-based model. It is a Modification of the algorithm derived from Google's PageRank and is based on eigenvector centrality measure and implements the concept of "voting" or "recommendation." The TextRank algorithms have relatively a stable accuracy than TF-IDF, and moreover, the precision rate is also better than TF-IDF. Due to the advantages of simple implementation, unsupervised and weak correlation of languages, the standard TextRank algorithm is suitable for single text and multi-text processing (Pan et al., 2019). The vital aspect of TextRank is that it gives an overall ranking sentence, whether it is concise summaries or longer, more explicative summaries, consisting of more than 100 words. TextRank is a graph-based algorithm that calculates the weight for a term. In this graph, a node is a term. If term A has a link with term B, it represents an edge from A to B. After constructing the whole graph, the Eq. (1) gives the weight of

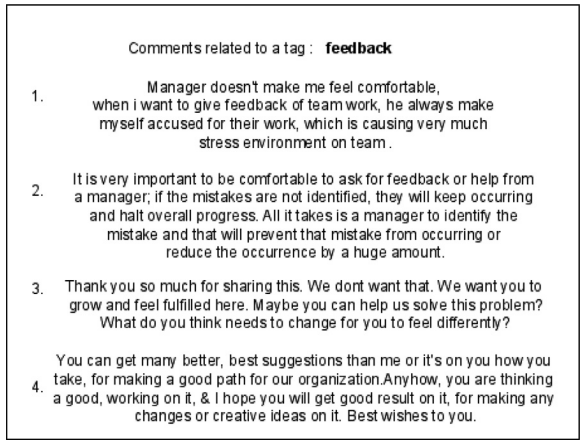

**Fig. 5.** Comments Related to Tag : feedback.

the term.

$$S(V_i) = (1 - d) + d * \sum_{j \in In(V_i)} \frac{1}{|Out(V_j)|} S(V_j) \qquad (1)$$

Where

- d is a damping factor, in case of no outgoing link it can be set between 0 to 1.
- $S(V_i)$ is weight of term i
- $In(V_i)$ inbound link of i
- $Out(V_j)$ Outgoing link of j
- $|Out(V_j)|$ No. of outbound links

## 5. Results

The output from the previous step produce tags for each group, and each group has multiple comments, so every comment in that group is assigned with a relevant tag. It may be possible that the comment does not have that keyword, but it is relative to the comment and gives the sense of the comment. Once the relevant keywords in the form of tags for the pulse comments are obtained, it is easy to find out the insights. It entirely depends on the use cases, and the applications could be limitless.

### 5.1. Keyword based comments retrieving

The study provides the logistics leaders with comments about a particular tag or keyword that they are more concerned with, such as harassment, life, management, attitude, and similar. It provides an opportunity for them to get what employees are talking about and how they can take appropriate actions to improve employee engagementfor example, the comments related to feedback (Refer Fig. 5). The Fig. 5 reveals the comments for the keyword "feedback" and signifies the organization's feedback system. It provides the management with actionable insight to identify the radical cause of disengagement due to the feedback management system and breakthrough to eliminate it. Similarly, logistics leaders can identify the comments related to other tags and determine the cause.

### 5.2. Identifiedkeywords combination

The study also provides the logistics leaders with comments that contain a particular combination of keywords. For example, they are more interested in getting the comments related to a combination of keywords like work-life, satisfaction, manager, or even they may be interested in the combination of tags or keywords like harassment, manager, office. The top three and least three occurring combinations in terms of their occurrence are (Refer Fig. 6). The results in Fig. 6 show that most employees are concerned about the office environment, particularly for conversation among team members. The least three combinations show that employees efficiently balance work-life and time, and very few employees desire an increase in office attitude. These combinations of keywords will give the gist of office feedback and provide directions to management for taking proper actions to maintain office pulse.

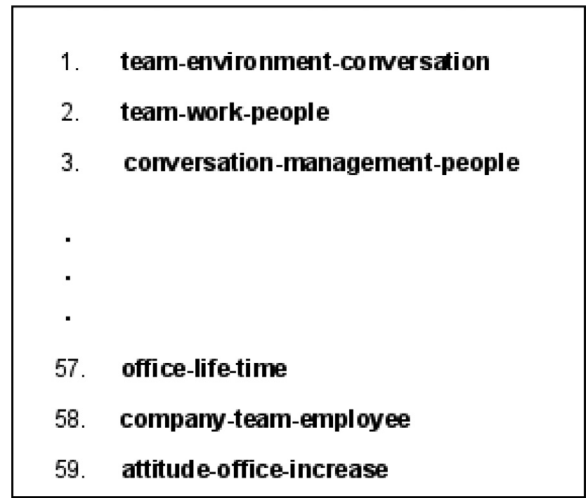

**Fig. 6.** The Most(1–3)and Least(57–59) frequent identified tags/keywords combinations.

### 5.3. Most frequent identified tags/keywords

The commonly used word provides insights about the frequency of using it and words that can potentially help management take action. For example, management can get insight by identifying the most frequently identified tag. It gives them a sense of employee's feeling in the organization (Refer Fig. 7).

## 6. Discussion

### 6.1. Theoretical contributions

This study provides an overview of the employee engagement theory. The philosophy of employee engagement is not a universal measure. Instead, it is a blend of interdisciplinary experience and analysis that addresses the customer's motivational needs and the organizational processes to reach broader organizational goals.

The study shows that staff commitment is a two-way street: a mutual partnership of trust and respect between employer and employee. It needs logistics leaders to explicitly and thoroughly convey their objectives with staff, inspire employees at their skill levels, and develop a working atmosphere and organizational culture where commitment will prosper. At best, engagement is a sign of growth. Engaged employees are more likely to invest in their jobs, contributing to better working efficiency and more significant revenue.

The supply chain industry is on the edge of big data analytics and predictive analysis, but issues such as information sharing and minimizing environmental uncertainties remain untouched (Wamba et al., 2018). The implementation of the information sharing system does not have a significant impact on engagement (Li and Sandino, 2018). Furthermore, the supply chain's decision support system primarily addresses supplier concerns and delivery, distribution, and transport (Teniwut and Hasyim, 2020). There is an increasing range of methods available for employee engagement. However, all of these methods appear to complicate the problems, fail to focus on what is essential, and are burdensome for those concerned, so they do not produce lasting results.

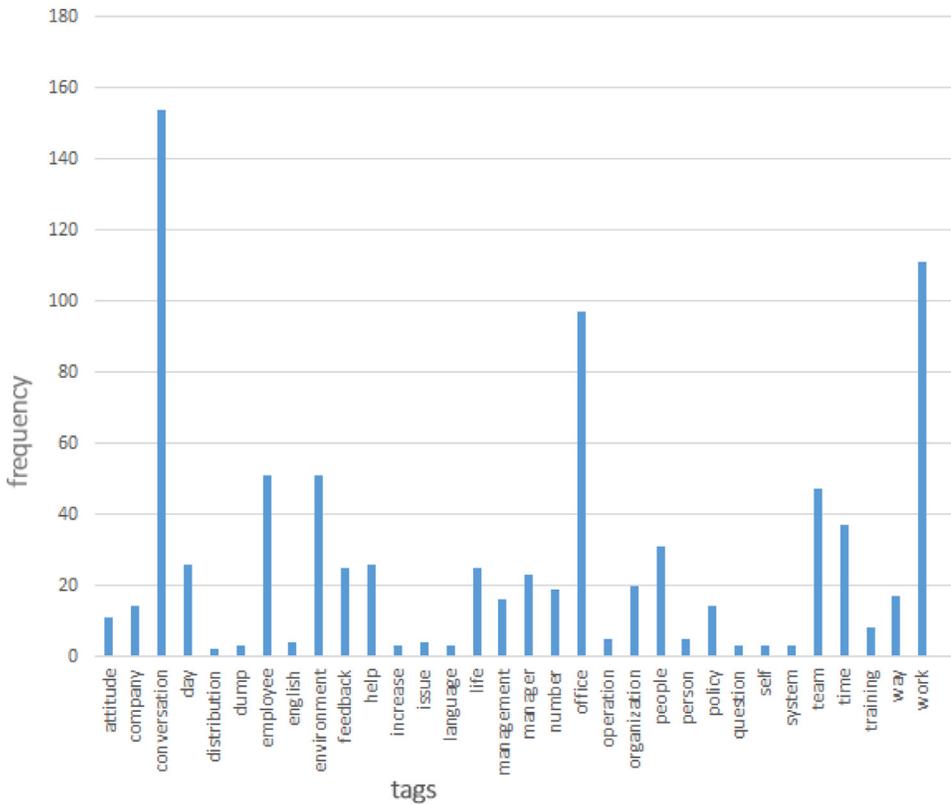

**Fig. 7.** The frequency of identified tags.

These various challenges for the sustainability of the supply chain include lack of digital culture, weak research and development, and the unavailability of any shared ground to develop an intelligent system (Luthra and Mangla, 2018). Different versions of engagement surveys cover the various dimensions of engagement. The challenges posed by existing engagement surveys within a multinational company mean a clear need for technical progress to improve employee engagement assessment and analysis (Burnett and Lisk, 2019; Sgarbossa et al., 2020). The work that enhances involvement by getting insights from the involved process must be centered, and research from leaders' and managers' perspectives should also be carried out to increase employee participation in the company (Bakker and Demerouti, 2008; Tran, 2018). The digital revolution has been brought on by the evolution of information and communication technologies (ICT). ICT facilitates the fourth industrial revolution known as Industry 4.0. The technology policies of organizations concerning employees have a good impact on their digital supply chain capacity, thereby improving supply chain efficiency (Queiroz et al., 2019). Supply chain processes combined with advanced information technology-driven industry 4.0 contribute to a sustainable supply chain culture (Luthra and Mangla, 2018). The logistics 4.0 definition involves coordination between computational mechanisms and ongoing processes. Smartness, correctness, and responsiveness are the key characteristics of this revolution (Monostori, 2018).

In extension to current literature, this study aims to implement AI's technological advancements by using NLP to summarize and provide a gist of employee feedback by identifying actionable insights that enhance employee engagement in a logistics organization.

### 6.2. Implications for Practice

Engagement does not take place in the imagination. It requires to be anchoring in the culture of a company. A corporate culture briefly describes the entirety of a company's activities, decisions, and management behavior. A leader must understand corporate culture and must change the corporate culture to match the strategic objectives. Creating a culture of employee engagement involves "checking with their employees to ensure that the organization's purpose is in line with existing working practices and ways of doing things."

Companies that communicate clearly and accurately can create trust among employees quickly. In short, management needs to know the fundamentals of empowering the workforce and providing a way to calculate the decisions' outcomes and make a commitment. Listen to the team on their terms and conditions and allow them to provide feedback and suggestions confidentially and anonymously. They will hear their frank opinions this way, not just what they think management wants to hear.

This study aims to enhance or bring employee engagement to perfection in a logistics company. The study will also identify the quantifiable insights from feedback comments and use them to identify some actions that logistics leaders can take to create change; hence they can better the scope of employee engagement. It will give them the momentum to pursue long-term action plans and to take action to secure fast gains. The research will also assist them in the following respects.

- Invest time in finding out what to add to the list to improve the scope of employee engagement
- Concentrating on development opportunities
- Delivering responsive leadership
- Focus on wellness strategies to improve work-life balance
- Setup exciting office culture by creating an office environment more compelling
- Identifying the most important sections to participate in maximizing employee engagement

### 6.3. Futureresearch directions

The commitment of employees is an essential factor in strengthening the reputation of the organization and its stakeholders. Today, an en-gaged workforce in the dynamic market has proved to be a key source of business success. This study focuses on identifying quantifiable insights that enable supply chain leaders to take adequate measures to balance trust and engagement and thus create an engaged workforce. This study's results are based on the data; only the pulse survey comments have been analyzed. Future research on employee engagement in logistics can be conducted to analyze employee social media content to identify employee feelings and monitor behavior trends. Authors, therefore, feel that NLP incorporation in the social media study can also contribute to the workforce engagement program and assist organizations in rationalizing the returns of the program. Furthermore, it will be interesting to create interactive dashboards that show a systematic division of the well-being of an enterprise, recognize areas that need consideration and allow users to go into data as deeply as they wish.

## 7. Conclusion

In recent years, pulse surveys have become increasingly common, particularly in the context of staff engagement programs. Literature shows that today many companies are searching for the next big thing. They conduct annual, monthly, or weekly surveys to find out what they should do for effective office vibes. However, these surveys provide essential details, likewise the annual or monthly report, but never be instruments that support logistics managers, and employees in their everyday work. Thus, i-Pulse commitment appears to be to carry the progress of the work performed on employee engagement to a higher level. This AI-enabled solution provides profound insights into employee feelings through natural language text analysis and helps logistics leaders understand their employees' concerns. It also obtains hands-on insights to identify the cause of disengagement in an organization and breakthrough to eliminate it. This study reveals insights for building organizational value to achieve a competitive advantage in the logistics field. This insight offers the expression of employees and allows logistics managers to consider the degree to which employees believe their management or organization. i-Pulse assists the management to create an outline that address the expectations of employees. As engagement is a key to achievement and sustainability, these insights enable management to build personalized strategies and boost employee work-life balance. To have a meaningful and beneficial impact on observable business performance, the i-Pulse needs to become part of the business. Such research can thus reveal untapped insights into increased employee engagement and participation in logistics activities.

**Declaration of Competing Interest**

The authors declares that they have no conflict of interests to the work reported in this paper.

**Acknowledgment**

The authors would like to thank the editors and anonymous reviewers for providing insightful suggestions and comments to improve the quality of research paper.